\documentstyle[11pt,newpasp,twoside,epsf]{article}

\begin{document}

\title{Confirming EIS Clusters\\Optical and Infrared Imaging}
\author{L.F. Olsen, H.E. J{\o}rgensen}
\affil{Astronomical Observatory, Juliane Maries Vej 30, DK-2100 Copenhagen, Denmark}
\author{M. Scodeggio, L. da Costa, R. Rengelink}
\affil{ESO, Karl-Schwarzschild-Str. 2, D-85748 Garching b. M\"{u}nchen, Germany}
\author{M. Nonino, A. Biviano, M. Ramella, W. Boschin}
\affil{Osservatorio Astronomico di Trieste, Via G.B. Tiepolo 11, I-31444 Trieste, Italy}

\begin{abstract}
Clusters of galaxies are important targets in observationally cosmology,
as they can be used both to study the evolution of the galaxies 
themselves and to constrain cosmological parameters. Here we 
report on the first results of a major effort to build up a sample of 
distant galaxy clusters to form the basis for further studies within 
those fields. We search for simultaneous overdensities in color and space to
obtain supporting evidence for the reality of the clusters.
We find a confirmation rate for EIS clusters of 
$66\%$, suggesting that a total of about 80 clusters with $z\geq 0.6$ are within 
reach using the EIS cluster candidates. 
\end{abstract}


\section{Introduction}

Distant massive clusters of galaxies are very important targets in
observational cosmology, as they can be used both for the study of
galaxy populations and to derive strong constraints on the parameters
of the cosmological models still debated. It is generally
recognized that the most straightforward method to identify
high-redshift \( (z>0.4) \) clusters is based on the detection of
their diffuse X-ray emission, but this method becomes prohibitively
expensive as the cluster redshift increases above \( z\simeq 0.8 \) to
be applied to extensive searches of such objects (ROSAT PSPC
exposures of \( \simeq 60 \) ksec are needed to detect clusters at \(
z\simeq 1). \) On the contrary, with the advent of large-field CCD
cameras, it is now possible to carry out large-area optical imaging
surveys to the depth required to identify rich clusters at these high
redshifts using relatively limited amounts of telescope time.
Preliminary efforts towards this goal, although still limited to
relatively small areas, include the Palomar Distant Cluster Survey
(Postman et al. 1996) and the ESO Imaging Survey (EIS, Nonino et
al. 1999).
 
Distant cluster candidate samples identified from imaging data could
however suffer from severe contamination effects, most likely caused
by projections along the line of sight, or large-scale fluctuations in
the galaxy distribution, that could make the final spectroscopic
confirmation process very expensive in terms of telescope time
requirements. In this paper and the related paper by Biviano et al.
(this volume) we report on a major effort towards quantifying the 
efficiency of optical searches as well as improving the efficiency 
of the spectroscopic follow up. Our starting point is the
sample of high-redshift cluster candidates identified from the \( I
\)-band EIS data over \( \simeq 14 \) deg\( ^{2} \) (Olsen et
al. 1999a,b; Scodeggio et al. 1999). This sample is composed of 302
cluster candidates, with estimated redshifts in the interval \(
0.2\leq z\leq 1.3 \) (median estimated redshift 0.5), that were
identified using the matched-filter algorithm described by Postman et
al. (1996).

To effectively address the question of how one efficiently produces a
sample of confirmed clusters from the cluster candidates we are
pursuing two different follow-up strategies. One is to carry out
further deep optical or near-infrared imaging observations, aimed at
the detection of the red sequence of cluster early-type galaxies in
color-magnitude diagrams. In this case the final spectroscopic
observations will be carried out only for those candidates where such
a sequence has been detected, targeting preferentially the galaxies in
the sequence. The second strategy eliminates this intermediate
observations, and is based on multi-object ``blind'' spectroscopic
observations of the cluster candidates (see Biviano et al, this 
volume). In this paper we report on the results of the first set of
deep pointed imaging observations, carried out at various telescopes to 
obtain $V$, $J$ and $K_s$ photometry for the distant clusters. We have 
obtained data for 27 EIS candidates and investigate the $(V-I)\times I$,
$(I-K_s)\times K_s$ and $(J-K_s)\times K_s$ color-magnitude diagrams.

\section{Observations and Data Reduction}

The data considered in this work consist primarily of pointed infrared and
optical observations obtained at different telescopes,
combined with the EIS $I$-band data.
The pointed observations include 1) 15 candidates observed
in $J$ and $K_s$ using SOFI at the ESO 3.5m New Technology Telescope (NTT)
at La Silla; 2) 4 candidates observed at the Danish 1.5m telesope at La
Silla; and 3) 13 candidates observed at the 2.5m Nordic Optical
Telescope at La Palma. Table~\ref{tab:inst} summarizes the
available data; giving 1) the instrument; 2) the
detector type; 3) the pixel scale in arcsec/pixel; 4) the field of view
in arcmin; 5) the typical limiting magnitude for the object catalogs;
and 6) the number of candidates observed with that instrument.

\begin{table}
\caption{Instrument characteristics}
\label{tab:inst}
\begin{tabular}{llrrlr}
\hline\hline
Instrument & CCD & pix.scale & field-of-view & $m_{lim}$ & cand.\\
\hline
SOFI & Rockwell, Hawaii & 0.29 & 4.9'$\times$4.9' & $J\sim22.5$ & 15 \\
@NTT & 1024$\times$1024 & & & $K_s\sim 20$ & 15 \\
DFOSC & Loral/Lesser & 0.39 & 13.7'$\times$13.7' & $V\sim 25$ & 4\\
@D1.5m& 2052$\times$2052\\
ALFOSC & Loral/Lesser & 0.19 & 6.4' $\times$6.4' & $V\sim 25.5$ & 13 \\
@NOT& 2052$\times$2052\\
\hline
\hline
\end{tabular}
\end{table}

All the data were reduced using the pipeline software developed for the
EIS project (Nonino et al., 1999; da Costa et al., 1998).
Optical images were reduced using standard
IRAF tasks, while the astrometric and photometric calibrations were
performed using specially developed pipeline programs. The
infrared images were reduced using the Eclipse software package
(Devillard, 1998) and the calibrations were based on the same programs
as for the optical data.

Color catalogs are derived from the
single band galaxy catalogs obtained using SExtractor (Bertin \& Arnouts, 1996). 
A procedure in the EIS pipeline combines the separate object detections into 
a color catalog, based on positional match between
the single band detections.  
Colors are computed from total magnitudes (SExtractor mag\_auto,  
see also Prandoni et al, 1999). 
Only objects
with SExtractor stellarity index $\leq0.8$ and extraction 
flags $\leq 3$ are included in the single band catalogs.

\section{Color confirmation}

\begin{figure} 
\plotfiddle{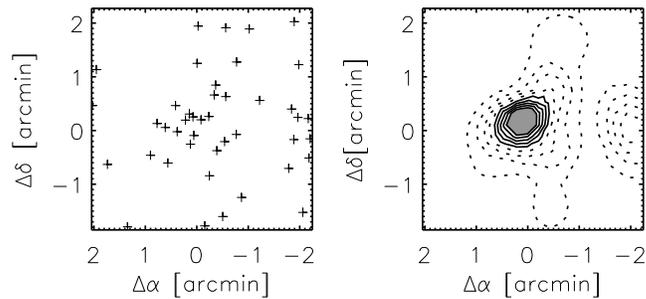}{3.2cm}{0}{110}{110}{-130}{-10}
\caption{Example of a color slice containing a confirmed cluster. The 
left panel is the actual galaxy distribution at that particular color
and the right panel shows the corresponding density map with the 
shaded region marking the $95\%$ confidence region.}
\label{fig:slices} 
\end{figure}

The EIS cluster search was based on $I$-band data only, using the
characteristic luminosity function and spatial distribution of the
cluster galaxies as the basis for the adopted filter 
(see Olsen et al., 1999a). 
The confirmation process is based on two
well known properties of the cluster early-type galaxy population,
namely its preferential location in the core of rich clusters, and
the small scatter shown by its colors. Therefore our
confirmation procedure is constructed to detect simultaneous color 
and space overdensities.

To explore in detail such three-dimensional space 
we split the galaxy color catalogs in slices of color 
0.3mag wide (the width is dictated by the estimated error of the faintest galaxies). We make two such
sets of slices shifted by half a bin width
to assure an equally good coverage at all
colors. For each color slice we smooth the galaxy distribution to
obtain the density in a pixel grid with a typical scale of 0.25arcmin.
To assess the significance of the detected overdensities we use
500 simulations for each color slice and thereby compute the density
distribution expected from a uniform galaxy distribution. We consider
a cluster confirmed when a density peak with a confidence level 
$\geq95\%$ is detected within 1arcmin from the original catalog position.
In figure~\ref{fig:slices} we show an example of the
galaxy distribution and corresponding density maps for a
confirmed cluster, the shaded area corresponding to the region where
a confidence level above 95\% is reached.

\begin{figure}
\plotfiddle{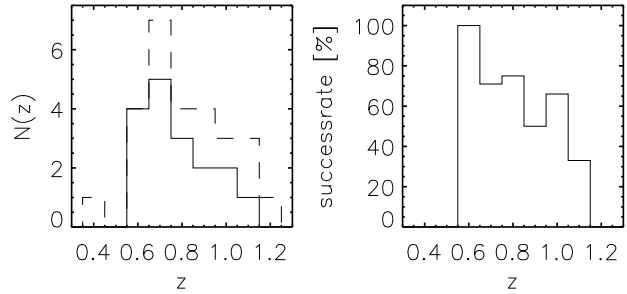}{3cm}{0}{110}{110}{-130}{-10}
\caption{Left panel: The number of observed clusters as function of
estimated redshift (dashed line) and the number of confirmed clusters
(solid line). Right panel: Confirmation rate as a function of redshift
for the 27 EIS cluster candidates.}
\label{fig:conf_rate}
\end{figure}

The method was applied to all candidates for which deep pointed 
observations were available and in figure~\ref{fig:conf_rate} we show
the redshift dependence of the confirmation rate. It is seen that the 
confirmation rate is about $66\%$ in the entire redshift interval, 
suggesting that a sample of 80 clusters at $z\geq0.6$ is within 
reach, based on the EIS cluster catalog. As discussed by 
Biviano et al. (this volume) the confirmation rate derived from
spectroscopic observations supports this result.

\section{Conclusions} 

Using color information we obtain strong evidence that 18 out 
of 27 ($\sim66\%$) of the EIS cluster candidates analyzed in this work are physical 
associations. Their estimated redshifts 
extend to about $z\sim1.0$. If the color confirmed
clusters are spectroscopically confirmed this method opens the 
possibility of obtaining a sample of 80 clusters of galaxies at
$z\geq0.6$ within very short timescales. This cluster sample will
be invaluable to any study of high redshift clusters and of the
cluster galaxy populations. 

\end{document}